\documentclass[10pt,conference]{IEEEtran}
\IEEEoverridecommandlockouts
\usepackage{cite}
\usepackage{amsmath,amssymb,amsfonts}
\usepackage{algorithmic}
\usepackage{graphicx}
\usepackage{textcomp}
\usepackage{xcolor}
\usepackage{amsthm}
\usepackage{subfigure}
\usepackage{colortbl}
\usepackage{soul}  
\usepackage{enumitem}
\usepackage{listings}
\usepackage{xcolor}
\usepackage{fancyhdr}
\definecolor{skyblue}{rgb}{0., 0.72, 0.92}
\usepackage{amsmath, amsfonts}
\usepackage[outline]{contour}
\usepackage[linesnumbered,ruled,vlined]{algorithm2e}
\SetKwInOut{Input}{Input}
\SetKwInOut{Output}{Output}

\SetCommentSty{mycommfont}

\usepackage{multirow}
\usepackage{booktabs}
\usepackage{balance}

\usepackage[utf8]{inputenc} 
\usepackage[T1]{fontenc}    
\usepackage{hyperref}       
\usepackage{url}            
\usepackage{booktabs}       
\usepackage{amsfonts}       
\usepackage{nicefrac}       
\usepackage{microtype}      
\usepackage{multirow}
\usepackage{rotating}
\usepackage{bbding}
\usepackage{subfigure}
\usepackage{amssymb}
\usepackage{listings}
\usepackage{enumitem}
\hypersetup{
    colorlinks=true,
    linkcolor=blue,
    filecolor=magenta,      
    urlcolor=blue,
    pdftitle={Overleaf Example},
    pdfpagemode=FullScreen,
    }
\usepackage{amsmath}

\theoremstyle{definition}

\definecolor{lightgray}{rgb}{0.78, 0.78, 0.78}
\sethlcolor{lightgray}  
\definecolor{answercolor}{RGB}{240, 240, 240}

\def\BibTeX{{\rm B\kern-.05em{\sc i\kern-.025em b}\kern-.08em
    T\kern-.1667em\lower.7ex\hbox{E}\kern-.125emX}}
\begin{document}


\title{CodeS: Towards Code Model Generalization Under Distribution Shift}


\DeclareRobustCommand{\IEEEauthorrefmark}[1]{\smash{\textsuperscript{\footnotesize #1}}}

\author{\IEEEauthorblockN{Qiang Hu\IEEEauthorrefmark{1*},
Yuejun Guo\thanks{*Equal Contribution.}\IEEEauthorrefmark{2*}, 
Xiaofei Xie\IEEEauthorrefmark{3}, 
Maxime Cordy\IEEEauthorrefmark{1},
Mike Papadakis\IEEEauthorrefmark{1},
Lei Ma\IEEEauthorrefmark{4,}\IEEEauthorrefmark{5} and
Yves Le Traon\IEEEauthorrefmark{1}}
\IEEEauthorblockA{\IEEEauthorrefmark{1}University of Luxembourg, Luxembourg\\
\IEEEauthorrefmark{2}Luxembourg Institute of Science and Technology, Luxembourg \\
\IEEEauthorrefmark{3}Singapore Management University, Singapore  \\
\IEEEauthorrefmark{4}University of Alberta, Canada
\quad
\IEEEauthorrefmark{5}The University of Tokyo, Japan
}}


\maketitle
\thispagestyle{fancy}
\pagestyle{fancy}
\cfoot{\thepage}
\renewcommand{\headrulewidth}{0pt} 
\renewcommand{\footrulewidth}{0pt}

\begin{abstract}

Distribution shift has been a longstanding challenge for the reliable deployment of deep learning (DL) models due to unexpected accuracy degradation. Although DL has been becoming a driving force for large-scale source code analysis in the big code era, limited progress has been made on distribution shift analysis and benchmarking for source code tasks. To fill this gap, this paper initiates to propose CodeS, a distribution shift benchmark dataset, for source code learning. Specifically, CodeS supports two programming languages (Java and Python) and five shift types (task, programmer, time-stamp, token, and concrete syntax tree). Extensive experiments based on CodeS reveal that 1) out-of-distribution detectors from other domains (e.g., computer vision) do not generalize to source code, 2) all code classification models suffer from distribution shifts, 3) representation-based shifts have a higher impact on the model than others, and 4) pre-trained bimodal models are relatively more resistant to distribution shifts. 
\end{abstract}


\section{Introduction}
\label{sec:intro}
The rapid advance of open source platforms (e.g., GitHub~\cite{githubWeb}), as well as a large number of (both publicly and industry internally) available software systems, brings challenges to traditional software analysis techniques, but offers new opportunities in understanding and reusing existing ``big code'' for software development by the data-driven approach. In particular, to enable the large volume of source code analysis, there comes a new trend in both software engineering and machine learning communities -- \emph{machine learning for source code}~\cite{allamanis2018survey,survey2022dl} in the last decade where deep neural networks (DNNs) have been widely employed for source code learning and achieved remarkable success in various tasks, e.g., code summarization~\cite{alon2018code2seq}, method naming~\cite{alon2019code2vec}, and source code classification~\cite{mou2016convolutional}. 

In practice, the main activity of \emph{machine learning for source code} remains at the stage of designing effective code representation techniques and model architectures~\cite{guo2021graphcodebert,alon2019code2vec}, which are evaluated on the \emph{pure} test data. Specifically, the most commonly used evaluation methodology of existing works is simply splitting a dataset into training, validation, and test sets and testing a model on the test set. In this way, the reported performance on the test set can only reflect the model performance on in-distribution data since the test set follows a similar distribution as the training set. However, with the rapid evolution of technology and software, new source code is programmed every day. Distribution shift often happens in the new data after the model has been deployed in the wild, which becomes a big challenge, and causes the model to produce unreliable predictions~\cite{wilds2021}. Here, distribution shift means that the test set follows a different data distribution from the training set. For example, in code learning models, compared to the training data, the new code data might be written by new developers (with different programming styles or habits) or collected from new projects using different techniques. Therefore, datasets with different types of distribution shifts are needed for the evaluation of code models.


In domains like computer vision (CV)~\cite{cifar10c2019dan} and natural language processing (NLP)~\cite{wilds2021}, data distribution shift has already gained considerable attention and multiple benchmarks have been constructed. For example, CIFAR-10-C~\cite{cifar10c2019dan} provides 15 types of algorithmically generated corruptions that generate images with different distributions from CIFAR-10. Wilds~\cite{wilds2021} collects natural datasets with distribution shifts including both image and text data in the wild. However, the study in the source code domain is still at a very early stage. Although some distribution shifts on source code~\cite{wilds2021,li2021estimating} are defined, they only considered rather simple scenarios, e.g., shifts from cross projects or time periods. The fine-grained and important features of source code data, for example, the frequency of represented tokens, are missed. Besides, existing datasets are not multi-language friendly because they only support one specific programming language such as Java or Python. The above issues and challenges limit the use of these datasets and the generality of conclusions that were drawn.

In this paper, we introduce CodeS, a distribution shift benchmark dataset to support the new evaluation paradigm for code models -- generalization ability evaluation. Overall, CodeS covers 5 types of code-level distribution shifts from 2 perspectives, natural and representation-based distribution shifts. Here, the representation-based distribution shifts can measure the lexical difference between two programs. Compared to natural distribution shifts, it is more fine-grained and controllable, and thus can be used to quantify the generalization ability of models more precisely. For natural distribution shift, we introduce \emph{cross task shift, cross programmer shift}, and \emph{time period shift}. For the representation-based one, we define \emph{token frequency shift and concrete syntax tree (CST) similarity shift}. Specifically, to build the datasets of CodeS, we first collect 3 in-distribution sets: two (Java250-S and Python800-S) are collected from existing code classification datasets (Java250 and Python800~\cite{codenet2021}), and one (Python75) is crawled from the AtCoder site~\cite{atcoder2012} by ourselves. Afterward, for each set, we generate 5 versions of data shifts. Finally, CodeS contains 16 groups of in/out-of-distribution (ID/OOD) code data written in 2 programming languages, Java and Python.

Based on the constructed datasets, we conduct experiments to investigate 1) the effectiveness of existing OOD detectors in distinguishing ID and OOD source code data, 2) the rationality of the definitions of source code distribution shift, and 3) the generalization ability of existing source code models on classification tasks. For the rationality exploration, we employ existing OOD detectors to quantify the shift degree. Generally, the more different the two sets are, the easier the OOD detector can distinguish them. The experimental results demonstrated that the well-designed OOD detectors for CV and NLP failed to distinguish the OOD code, and only the simple softmax score-based OOD detector performs well. The representation-based distribution shifts have a higher impact on the performance of the models, among which the token distribution shift brings the biggest accuracy degradation. 

To summarize, the main contributions of this paper are:

\begin{itemize}[leftmargin=*]
\itemsep0em
\item We propose CodeS\footnote{Please visit our companion site for more details (data, model, implementation, and results): \url{https://sites.google.com/view/codes-distribution-shift}.}, a benchmark dataset that provides fine-grained distribution shift datasets to support the new evaluation paradigm (generalization ability evaluation) for code models. This is the first benchmark that covers two programming languages (Java and Python) and representation-based distribution shifts.  
\item We perform a comprehensive evaluation of the effectiveness of existing OOD detectors and the rationality of the defined distribution shifts. 
\item We conduct an initial investigation on popular source code classification models using CodeS and demonstrate that code representation-based (token and CST) distribution shifts cause significant degradation in model performance. 
\end{itemize}

\section{Related Work}
\label{sec:bac}
\paragraph{Source code learning} Since source code can be represented as text data (e.g., sequence of tokens) and structural data (e.g., data flow), deep neural networks have been employed for learning such representations to solve different tasks in recent years~\cite{allamanis2018survey}. In which, source code classification is one of the widely studied tasks for code understanding, and many source code classification applications have been proposed.~\cite{alon2019code2vec} proposed the very early work, Code2Vec, which aims at predicting the function name of a code snippet by learning the distributed code representations.~\cite{lu2021codexglue} proposed a large-scale dataset CodeXGLUE for code understanding. It supports 2 code binary classification tasks that are related to the privacy and vulnerability of code, clone detection, and defect detection. More recently, researchers from IBM~\cite{codenet2021} proposed CodeNet which contains source code solution classification tasks with multiple programming languages, e.g., Java and Python. Different from existing works that only consider clean datasets without data shift, we construct the dataset including both clean data and distribution shift data, which can be better used to measure the performance of trained models from a more practical perspective.    

\paragraph{Distribution shift in source code} Some works have studied the distribution shift in source code data. Table~\ref{tab:comparison} lists the overview of the comparison between existing datasets and CodeS. More detailed, Wilds, proposed by~\cite{wilds2021}, is the first work that mentioned the data distribution shift problem for code completion in source code learning. Wilds first defined the cross-project distribution shift that the shift might come from the data biases of different code repositories. Besides,~\cite{li2021estimating} conducted an empirical study to investigate the prediction uncertainty of models under three types of distribution shifts, code collected from different projects, code written by different programmers, and code collected across different time periods. The authors found that the uncertainty metrics proposed in the computer vision domain are not fully applicable for source code tasks, e.g., code summarization and code completion. Moreover, in the latest work,~\cite{nie2022impact} discussed that the trained models should be evaluated in different methodologies according to the real use cases in the code summarization task. For example, in the evolution scenario, the developers should consider if the model trained at time $t_{0}$ can still be used at time $t_{1}$ in the future, which can be another data distribution shift. Compared to the above works, CodeS provides datasets with more comprehensive and fine-grained distribution shifts, i.e., more programming languages and more diverse data shifts. Furthermore, based on the benchmarks, we evaluate the rationality of our distribution shift definitions using OOD detectors. 

\begin{table}[!t]
\caption{Comparison between our developed CodeS and existing datasets.}
\label{tab:comparison}
\centering
\resizebox{.5\textwidth}{!}{
\begin{tabular}{lcccccc}
\hline
 & \textbf{Programming Language} & \textbf{Task} & \textbf{Programmer} & \textbf{Time-stamp} & \textbf{Token} & \textbf{CST} \\ \hline
\textbf{Wilds}~\cite{wilds2021} & Python & \Checkmark & \XSolid & \XSolid & \XSolid & \XSolid \\
\cite{li2021estimating} & Java & \Checkmark & \Checkmark & \Checkmark & \XSolid & \XSolid \\
\cite{nie2022impact} & Java & \Checkmark & \XSolid & \Checkmark & \XSolid & \XSolid \\
\textbf{CodeS} & Python, Java & \Checkmark & \Checkmark & \Checkmark & \Checkmark & \Checkmark \\ \hline
\end{tabular}
}
\end{table}

\section{The CodeS Dataset}
\label{sec:shifts}

\subsection{Dataset construction}
\label{subsec:configurations}
\textbf{Distribution shift definition.} We design five types of distribution shifts for source code from two perspectives: natural distribution shift and representation-based distribution shift. Natural distribution shift comes from common code style changes that a model could face every day, including the change of task, time period, and programmer. On the other hand, since the source code is often transformed to different representations before feeding to a model, such as the linear sequence of tokens and concrete syntax tree (CST)~\cite{codenet2021,Aroma2019}, we also consider distribution shifts based on the representation. Compared to the natural one, the fine-grained representation-based shift can reflect the implicit difference in data features. 

\emph{Task distribution shift:} ID data and OOD data target different tasks.

\emph{Programmer distribution shift:} ID data and OOD data target the same task but come from different programmers. Due to the programming habit, the distribution shift can happen across different programmers.

\emph{Time distribution shift:} ID code and OOD code target the same task but are written in different time periods. Assessing~\cite{Li2019ces} and maintaining~\cite{hu2022distribution} the model performance over time is critical to ensure the reliability and security of models given that the code can be modified or new users appear over time.

\emph{Token distribution shift: } ID code and OOD code target the same task but the frequencies of tokens that appear to them are different. In source code learning, transforming code into numeral vectors is fundamental to make it executable for deep learning models~\cite{survey2021}. A linear sequence of tokens is the typical and most important code representation, which is usually processed via tokenization, or lexical analysis. A token is the basic unit of the representation and can be a function name, an operator, or a punctuation sign. In practice, each token is represented by an integer to be compatible with models. Given the token sequences of two source code files, the straightforward difference is the appearance of tokens. 

\emph{Concrete Syntax Tree (CST) distribution shift: } ID code and OOD code target the same task but have different CST representations. The concrete syntax tree (also known as the parse tree or derivation tree) is another popular code representation, which includes the syntactic structure of code files. The difference between code files can be represented by the distance between CSTs.

\textbf{Raw data preparation.} The raw data of Python75 is collected from the AtCoder site~\cite{atcoder2012}, where new submissions and programmers are continuously announced in the contests. We selected the submissions from 75 tasks in 25 contests based on the following conditions: 1) the language should be Python3, 2) the tasks have at least 1,000 submissions, and 3) the status of the submission is accepted (AC). In total, we collected 200,462 programs for Python75. 
The raw data of Java250-S and Python800-S are from Java250 and Python800 provided by Project CodeNet~\cite{codenet2021}. In total, Java250-S and Python800-S contain 75000 and 240000 programs.

\textbf{Shift dataset creation.} For each type of distribution shift, we create a dataset consisting of the training set, ID test set, and OOD test set extracted from the raw data. The training and ID test sets share the same data distribution and both contribute to training a model. The OOD test set is used to evaluate the generalization ability of the trained model. Table~\ref{tab:size} shows the details of each dataset.

\begin{table}[h]
\caption{Data information (number of classes, data size per class).}
\centering
\label{tab:size}
\resizebox{.5\textwidth}{!}{
\begin{tabular}{lcccc}
\hline
\multirow{2}{*}{Data collection} & \multicolumn{4}{c}{Programmer, Time, Token, CST (Task)} \\ \cline{2-5} 
 & \#ID/OOD Classes & \#Training & \#ID test & \#OOD test \\ \hline
Python75 & 75/75 (65/10) & 732 (846) & 134 (154) & 134 (1000) \\
Java250-S & 250/250 (200/50) & 180 (225) & 60 (75) & 60 (300) \\
Python800-S & 800/800 (640/160) & 180 (225) & 60 (75) & 60 (300) \\ \hline
\end{tabular}
}
\end{table}

\emph{Task distribution shift.} We firstly randomly divide all tasks into the given number (\emph{\#ID Classes} and \emph{\#OOD Classes} in Table~\ref{tab:size}) of ID and OOD tasks. Next, in each ID task, we randomly select the number of \emph{\#Training} and \emph{\#ID test} code files as the training and ID test data, respectively. Finally, in each OOD task, a number of \emph{\#OOD test} code files are randomly selected as the OOD test data. 

\emph{Programmer distribution shift.} We follow the strategy used by~\cite{li2021estimating} to prepare this type of data. Specifically, for each task, we first randomly select specific 
programmers and consider their submissions as the OOD test data, the submissions from other programmers (at least two submissions) are added into training and ID test sets.  

\emph{Time distribution shift.} For each task, we sort the source code files according to the submission time and take the newest files as OOD test data\footnote{The time tag of Java250 and Python800 is unavailable, hence we do not create datasets with the time shift.}. The earlier code files are randomly split into the training and ID test sets, respectively. 

\emph{Token distribution shift.} Given a task, we first build a histogram of tokens using the token sequences of all code data (the bin number is equal to the total number of token types)~\cite{silverman1986book}. Then we take the token occurrence as a discriminator between ID and OOD data. For example, if some tokens are mostly/only used in certain code files, we consider these files to be either ID or OOD according to the file size and the pre-defined number of ID/OOD data. The ID set is further randomly split into the training and ID test sets. In this paper, we use the publicly available tokenizer tool provided by~\cite{codenet2021} to generate the token representations. 

\emph{CST distribution shift.} For each task, we calculate the average distance between each file and the others by the Robinson-Foulds distance between two CSTs. Here, we use the distance to refer to how far a code file shifts form the entire code set. The files with greater average distances are grouped into the OOD set, and the ones with smaller distances are grouped into the training and ID test sets. We use the parse tree generator provided by~\cite{codenet2021} to obtain the CST presentations. 

\section{Experimental setup}
\label{sec:exp}
\textbf{DNNs.} We consider 5 DNNs with different code representations. CNN (Sequence) is a model with both max and average global pooling operations (doublePoolClassDNN). MLP (Bag) is a DNN with dense layers (denseDNN). Both CNN (Sequence) and MLP (Bag) are provided by the Project CodeNet~\cite{codenet2021}. Besides, we apply 3 pre-trained bimodal models (RoBERTa~\cite{liu2020roberta}, CodeBERT~\cite{codebert2020}, and GraphCodeBERT~\cite{guo2021graphcodebert}) provided by the CodeXGLUE project~\cite{codexglue2021}. Each pre-trained model is fine-tuned using the training set. 

\textbf{OOD detectors.} An OOD detector is a method that, given a model, distinguishes ID and OOD data. Four widely used OOD detectors~\cite{fort2021exploring} are considered, Maximum Softmax Probability (MSP)~\cite{msp2017}, Out-of-Distribution detector for neural networks (ODIN)~\cite{odin2018enhancing}, Mahalanobis~\cite{ma18kim}, and Outlier Exposure (OE)~\cite{hendrycks2018deep}. The first three are output (softmax) score-based and do not require any change in the model. OE has a specific loss function and trains a new neural network using the same architecture as the given model. 

\textbf{Evaluation measures.} \emph{Accuracy} is a basic measure that calculates the percentage of data being correctly classified. Given a model, we measure its accuracy on ID and OOD test sets, respectively. Generally, a model will have greater accuracy on ID test data than on the OOD since its training data has the same distribution as the ID test set~\cite{wilds2021}. \emph{AUC-ROC} is short for the area under the receiver operating characteristics. The main difference between accuracy and AUC-ROC is that the latter does not depend on a specific classification threshold, hence it is a metric that computes better the inherent ability of the model to properly distinguish the different classes. An ideal OOD detector has the highest AUC-ROC score of 100\%.

\section{Results Analysis}
\label{sec:results}

\subsection{OOD detector and shift degree analysis}
\label{subsec:AUC-ROC}

\begin{table}[h]
\caption{AUC-ROC score and accuracy on different distribution shifts. Model: CNN (sequence). Average: mean over 4 OOD detectors.}
\label{tab:exp1-all}
\centering
\resizebox{.48\textwidth}{!}{
\begin{tabular}{lccccccc}
\hline
 & \multicolumn{5}{c}{AUC-ROC} & \multicolumn{2}{c}{Accuracy (\%)} \\ \cline{2-8} 
\multirow{-2}{*}{Shift type} & MSP & ODIN & Mahalanobis & OE & \cellcolor[HTML]{EFEFEF}Average & ID test & OOD test \\ \hline
\multicolumn{8}{c}{Python75} \\ \hline
Random & 50.16 & 62.83 & 49.07 & 49.64 & \cellcolor[HTML]{EFEFEF}52.93 & 96.91 & 97.01 (0.10 $\textcolor{blue}{\uparrow}$) \\
Task & 91.33 & 61.42 & 70.54 & 61.11 & \cellcolor[HTML]{EFEFEF}71.10 & 96.95 & 0.00 (0.00) \\
Programmer & 48.20 & 83.23 & 50.21 & 48.33 & \cellcolor[HTML]{EFEFEF}57.49 & 96.53 & 96.47 (0.06 $\textcolor{red}{\downarrow}$) \\
Time & 57.90 & 76.80 & 50.00 & 50.22 & \cellcolor[HTML]{EFEFEF}58.73 & 97.51 & 92.64 (4.87 $\textcolor{red}{\downarrow}$) \\
Token & 82.82 & 71.16 & 56.70 & 52.61 & \cellcolor[HTML]{EFEFEF}65.82 & 97.50 & 61.04 (36.46 $\textcolor{red}{\downarrow}$) \\
CST & 77.89 & 70.43 & 57.84 & 57.84 & \cellcolor[HTML]{EFEFEF}66.00 & 96.98 & 71.17 (25.81 $\textcolor{red}{\downarrow}$) \\ \hline
\multicolumn{8}{c}{Java250-S} \\ \hline
Random & 50.06 & 49.69 & 52.25 & 48.12 & \cellcolor[HTML]{EFEFEF}50.03 & 84.23 & 84.43 (0.20 $\textcolor{blue}{\uparrow}$) \\
Task & 80.99 & 44.85 & 70.78 & 52.67 & \cellcolor[HTML]{EFEFEF}62.32 & 87.89 & 0.00 (0.00) \\
Programmer & 53.72 & 49.36 & 46.33 & 49.22 & \cellcolor[HTML]{EFEFEF}49.66 & 85.86 & 82.84 (3.02 $\textcolor{red}{\downarrow}$) \\
Token & 68.65 & 52.28 & 49.47 & 52.63 & \cellcolor[HTML]{EFEFEF}55.76 & 89.01 & 65.85 (23.16 $\textcolor{red}{\downarrow}$) \\
CST & 60.32 & 41.85 & 50.98 & 51.53 & \cellcolor[HTML]{EFEFEF}51.17 & 86.69 & 75.91 (10.78 $\textcolor{red}{\downarrow}$) \\ \hline
\multicolumn{8}{c}{Python800-S} \\ \hline
Random & 49.67 & 37.63 & 48.55 & 50.00 & \cellcolor[HTML]{EFEFEF}46.46 & 77.91 & 78.40 (0.49 $\textcolor{blue}{\uparrow}$) \\
Task & 78.88 & 13.13 & 53.14 & 50.45 & \cellcolor[HTML]{EFEFEF}48.90 & 82.50 & 0.00 (0.00) \\
Programmer & 58.10 & 46.34 & 54.30 & 62.45 & \cellcolor[HTML]{EFEFEF}55.30 & 77.22 & 66.94 (10.28 $\textcolor{red}{\downarrow}$) \\
Token & 64.47 & 39.24 & 50.52 & 50.00 & \cellcolor[HTML]{EFEFEF}51.06 & 79.27 & 53.83 (25.44 $\textcolor{red}{\downarrow}$) \\
CST & 49.74 & 48.02 & 48.26 & 50.00 & \cellcolor[HTML]{EFEFEF}49.01 & 77.94 & 77.82 (0.12 $\textcolor{red}{\downarrow}$) \\ \hline
\end{tabular}
}
\end{table}

Since the data with task distribution shift is real OOD data (these data are not in the classification target of the model), we can use ID/OOD data split based on task distribution shift and random split to evaluate the performance of OOD detectors. A high AUC-ROC score obtained by the detector indicates that its discrimination ability is strong. Table~\ref{tab:exp1-all} shows the results, we can see that the simplest MSP detector produces the best result where the
difference between random and task distribution shifts is larger than by the well-designed Mahalanobis and OE. This phenomenon demonstrates that compared to image and text data, the two programs are more difficult to be distinguished. More powerful OOD detectors are needed. 

Then, to explore the reasonability of our distribution shift definitions, we check the shift degree caused by each definition. As shown in Table~\ref{tab:exp1-all}, regardless of the dataset, all types of distribution shifts (except for random splitting) cause performance degradation. Furthermore, we check the score of each definition. The results demonstrate that code representation-based (Token and
CST) distribution shifts cause greater performance degradation than the natural distribution shifts
(Programmer and Time) except for the task-based. 

\subsection{Model generalization analysis}
\label{subsec:embed}

\begin{table*}[ht]
\caption{Model accuracy (\%) on ID and OOD test sets given different DNNs and distribution shifts.}
\label{tab:exp3_emb}
\centering
\resizebox{.95\textwidth}{!}{
\begin{tabular}{lcccccccccc}
\hline
\multirow{2}{*}{Model} & \multicolumn{2}{c}{Random} & \multicolumn{2}{c}{Programmer} & \multicolumn{2}{c}{Time} & \multicolumn{2}{c}{Token} & \multicolumn{2}{c}{CST} \\ \cline{2-11} 
 & ID test & OOD test & ID test & OOD test & ID test & OOD test & ID test & OOD test & ID test & OOD test \\ \hline
\multicolumn{11}{c}{Python75} \\ \hline
CNN (Sequence) & 96.91 & 97.01 (0.1 $\textcolor{blue}{\uparrow}$) & 96.53 & 96.47 (0.06 $\textcolor{red}{\downarrow}$) & 97.51 & 92.64 (4.87 $\textcolor{red}{\downarrow}$) & 97.50 & 61.04 (36.46 $\textcolor{red}{\downarrow}$) & 96.98 & 71.17 (25.81 $\textcolor{red}{\downarrow}$) \\
MLP (Bag) & 92.22 & 92.07 (0.15 $\textcolor{red}{\downarrow}$) & 92.93 & 91.80 (1.13 $\textcolor{red}{\downarrow}$) & 92.35 & 88.13 (4.22 $\textcolor{red}{\downarrow}$) & 94.13 & 45.28 (48.85 $\textcolor{red}{\downarrow}$) & 93.53 & 63.45 (30.08 $\textcolor{red}{\downarrow}$) \\
RoBERTa & 98.31 & 98.13 (0.18 $\textcolor{red}{\downarrow}$) & 98.35 & 98.46 (0.11 $\textcolor{blue}{\uparrow}$) & 97.92 & 99.28 (1.36 $\textcolor{blue}{\uparrow}$) & 98.28 & 95.02 (3.26 $\textcolor{red}{\downarrow}$) & 99.47 & 86.65 (12.82 $\textcolor{red}{\downarrow}$) \\
CodeBERT & 98.37 & 98.29 (0.08 $\textcolor{red}{\downarrow}$) & 98.30 & 98.53 (0.23 $\textcolor{blue}{\uparrow}$) & 98.06 & 99.44 (1.38 $\textcolor{blue}{\uparrow}$) & 98.51 & 96.92 (1.59 $\textcolor{red}{\downarrow}$) & 99.52 & 87.04 (12.48 $\textcolor{red}{\downarrow}$) \\
GraphCodeBERT & 98.49 & 98.34 (0.15 $\textcolor{red}{\downarrow}$) & 98.40 & 98.63 (0.23 $\textcolor{blue}{\uparrow}$) & 98.11 & 99.52 (1.41 $\textcolor{blue}{\uparrow}$) & 98.50 & 96.37 (2.13 $\textcolor{red}{\downarrow}$) & 99.52 & 87.17 (12.35 $\textcolor{red}{\downarrow}$) \\ \hline
\multicolumn{11}{c}{Java250-S} \\ \hline
CNN (Sequence) & 84.23 & 84.43 (0.2 $\textcolor{blue}{\uparrow}$) & 85.86 & 82.84 (3.02 $\textcolor{red}{\downarrow}$) & - & - & 89.01 & 65.85 (23.16 $\textcolor{red}{\downarrow}$) & 86.69 & 75.91 (10.78 $\textcolor{red}{\downarrow}$) \\
MLP (Bag) & 70.58 & 71.61 (1.03 $\textcolor{blue}{\uparrow}$) & 72.40 & 71.59 (0.81 $\textcolor{red}{\downarrow}$) & - & - & 76.31 & 35.32 (40.99 $\textcolor{red}{\downarrow}$) & 76.03 & 50.86 (25.17 $\textcolor{red}{\downarrow}$) \\
RoBERTa & 94.87 & 95.26 (0.39 $\textcolor{blue}{\uparrow}$) & 95.53 & 95.04 (0.49 $\textcolor{red}{\downarrow}$) & - & - & 96.55 & 87.43 (9.12 $\textcolor{red}{\downarrow}$) & 96.79 & 87.47 (9.32 $\textcolor{red}{\downarrow}$) \\
CodeBERT & 95.67 & 96.10 (0.43 $\textcolor{blue}{\uparrow}$) & 96.29 & 96.25 (0.04 $\textcolor{red}{\downarrow}$) & - & - & 96.92 & 89.27 (7.65 $\textcolor{red}{\downarrow}$) & 97.41 & 88.63 (8.78 $\textcolor{red}{\downarrow}$) \\
GraphCodeBERT & 96.02 & 96.34 (0.32 $\textcolor{blue}{\uparrow}$) & 96.56 & 96.51 (0.05 $\textcolor{red}{\downarrow}$) & - & - & 97.29 & 89.89 (7.4 $\textcolor{red}{\downarrow}$) & 97.82 & 89.31 (8.51 $\textcolor{red}{\downarrow}$) \\ \hline
\multicolumn{11}{c}{Python800-S} \\ \hline
CNN (Sequence) & 77.91 & 78.40 (0.49 $\textcolor{blue}{\uparrow}$) & 77.23 & 66.94 (10.29 $\textcolor{red}{\downarrow}$) & - & - & 79.27 & 53.83 (25.44 $\textcolor{red}{\downarrow}$) & 77.94 & 77.82 (0.12 $\textcolor{red}{\downarrow}$) \\
MLP (Bag) & 66.90 & 67.00 (0.10 $\textcolor{blue}{\uparrow}$) & 67.29 & 63.39 (3.90 $\textcolor{red}{\downarrow}$) & - & - & 71.37 & 34.47 (36.90 $\textcolor{red}{\downarrow}$) & 66.39 & 67.77 (1.38 $\textcolor{blue}{\uparrow}$) \\
RoBERTa & 96.09 & 96.38 (0.29 $\textcolor{blue}{\uparrow}$) & 95.73 & 96.15 (0.42 $\textcolor{blue}{\uparrow}$) & - & - & 97.04 & 88.51 (8.53 $\textcolor{red}{\downarrow}$) & 95.72 & 97.31 (1.59 $\textcolor{blue}{\uparrow}$) \\
CodeBERT & 96.48 & 96.74 (0.26 $\textcolor{blue}{\uparrow}$) & 96.19 & 96.99 (0.80 $\textcolor{blue}{\uparrow}$) & - & - & 97.44 & 89.91 (7.53 $\textcolor{red}{\downarrow}$) & 96.19 & 97.77 (1.58 $\textcolor{blue}{\uparrow}$) \\
GraphCodeBERT & 96.69 & 96.89 (0.20 $\textcolor{blue}{\uparrow}$) & 96.26 & 97.24 (0.98 $\textcolor{blue}{\uparrow}$) & - & - & 96.19 & 97.77 (1.58 $\textcolor{blue}{\uparrow}$) & 96.42 & 97.98 (1.56 $\textcolor{blue}{\uparrow}$) \\ \hline
\end{tabular}
}
\end{table*}

Finally, we utilize CodeS to test the generalization ability of code models. Table~\ref{tab:exp3_emb} shows the results. Regardless of the dataset and model, the accuracy of the randomly split ID and OOD test sets are similar (the greatest difference is 1.03\% in Java250, MLP (Bag)). By comparing each model, we observe that the three pre-trained language models (RoBERTa, CodeBERT, and GraphCodeBERT) are more robust against the distribution shift than CNN (Sequence) and MLP (Bag). For example, in Python800-S with token distribution shift, all the pre-trained models can achieve more than 88\% accuracy on the OOD test set while CNN (Sequence) and MLP (Bag) only achieve by up to 53.43\% accuracy. This indicates that pretraining with diverse programming languages is helpful for producing models with better generalization ability on distribution shifted datasets. For instance, CodeBERT is pre-trained on natural language-programming language pairs in 6 programming languages (Python, Java, JavaScript, PHP, Ruby, and Go). In addition, by the accuracy drop on the OOD test sets, we can draw the same conclusion as the shift degree study that representation-based distribution shifts are more challenging.

\section{Future Plans}
In future work, we plan to:

\begin{itemize}[leftmargin=*]
\item design more practical types of shifts, e.g., shifts from the perspectives of software domains, coding styles, and programming languages. Besides, we will explore why existing OOD detection methods do not work well on CodeS and evaluate more methods from the SE community~(e.g., neuron coverage~\cite{ma2018deepgauge} and surprise adequacy~\cite{kim2019guiding}) that might detect OOD code data.
\item extend CodeS to support more code tasks, e.g., bug detection (by automatic bug injection)~\cite{bug2019li} and code summarization (by using the title of contests as the label).
\item build a larger distribution shift dataset based on our definitions from the industrial perspective (e.g., collect data from GitHub repositories).
\item investigate the practices (e.g., data augmentation) to improve code model generalization against OOD using CodeS.
\end{itemize}

\section{Conclusion}
\label{sec:conclu}
This paper proposed CodeS, the first distribution shift benchmark dataset that supports two programming languages and both natural and representation-based distribution shifts. We evaluated the effectiveness of existing OOD detectors, the shift degree introduced by different shift types, and the generalization ability of popular code classification models using CodeS. We found that the well-designed OOD detectors cannot generalize to code data and representation-based distribution shifts are more challenging than natural ones. CodeS pave the way for future study of source code learning, including OOD detection, model robustness enhancement, and more. 

\section*{Acknowledgments}
This work is supported by the Luxembourg National Research Funds (FNR) through CORE project C18/IS/12669767/STELLAR/LeTraon.
Lei Ma is supported in part by
JSPS KAKENHI Grant No.JP20H04168, 
JST-Mirai Program Grant No.JPMJMI20B8, as well as Canada CIFAR AI Chairs Program, and the Natural Sciences and Engineering Research Council of Canada (NSERC No.RGPIN-2021-02549, No.RGPAS-2021-00034, No.DGECR-2021-00019).

\bibliographystyle{IEEEtran}
\balance
\bibliography{IEEEabrv,sample-base}

\end{document}